\documentclass[twocolumn,tighten,times]{aastex63}
\newcommand\footnoterule{}
\def\footnotenew#1{
\skip\footins 12pt
\footnotesep 9pt
\renewcommand\thefootnote{\hskip-10pt \arabic{footnote}}
\renewcommand\footnoterule{\hskip-10pt \rule{\columnwidth}{0.5pt}}
\footnote{\leftskip-1.5mm \hspace{2mm} #1}
}

\graphicspath{{./}{figures/}}

\newcommand{\thetasl}{$\theta_\mathrm{SL}$}
\newcommand{\costheta}{cos$\,\theta_\mathrm{SL}$}

\defcitealias{2019ApJ...882...14M}{I}
\defcitealias{2019ApJ...887...59A}{II}

\accepted{for publication in ApJ}
\shorttitle{\small Origin of Spin--Orbit Alignment of Galaxy Pairs}
\shortauthors{\small Moon, An \& Yoon}
\begin{document}

\title{\large Living with Neighbors. III. The Origin of the Spin--Orbit Alignment of Galaxy Pairs: \\
A Neighbor Versus the Large-scale Structure}

\correspondingauthor{Suk-Jin Yoon}
\email{sjyoon0691@yonsei.ac.kr}

\author[0000-0001-7075-4156]{Jun-Sung Moon}
\affiliation{Department of Astronomy, Yonsei University, Seoul, 03722, Republic of Korea}
\affiliation{Center for Galaxy Evolution Research, Yonsei University, Seoul, 03722, Republic of Korea}

\author[0000-0003-3791-0860]{Sung-Ho An}
\affiliation{Department of Astronomy, Yonsei University, Seoul, 03722, Republic of Korea}
\affiliation{Center for Galaxy Evolution Research, Yonsei University, Seoul, 03722, Republic of Korea}

\author[0000-0002-1842-4325]{Suk-Jin Yoon}
\affiliation{Department of Astronomy, Yonsei University, Seoul, 03722, Republic of Korea}
\affiliation{Center for Galaxy Evolution Research, Yonsei University, Seoul, 03722, Republic of Korea}

\collaboration{3}{}

%% Note that the \and command from previous versions of AASTeX is now
%% depreciated in this version as it is no longer necessary. AASTeX 
%% automatically takes care of all commas and "and"s between authors names.

%% AASTeX 6.3 has the new \collaboration and \nocollaboration commands to
%% provide the collaboration status of a group of authors. These commands 
%% can be used either before or after the list of corresponding authors. The
%% argument for \collaboration is the collaboration identifier. Authors are
%% encouraged to surround collaboration identifiers with ()s. The 
%% \nocollaboration command takes no argument and exists to indicate that
%% the nearby authors are not part of surrounding collaborations.

%% Mark off the abstract in the ``abstract'' environment. 
\begin{abstract}
Recent observations revealed a coherence between the spin vector of a galaxy and the orbital motion of its neighbors. We refer to the phenomenon as ``the spin--orbit alignment (SOA)" and explore its physical origin via the IllustrisTNG simulation. This is the first study to utilize a cosmological hydrodynamic simulation to investigate the SOA of galaxy pairs. In particular, we identify paired galaxies at $z = 0$ having the nearest neighbor with mass ratios from 1/10 to 10 and calculate the spin--orbit angle for each pair. Our results are as follows. (a) There exists a clear preference for prograde orientations (i.e., SOA) for galaxy pairs, qualitatively consistent with observations. (b) The SOA is significant for both baryonic and dark matter spins, being the strongest for gas and the weakest for dark matter. (c) The SOA is stronger for less massive targets and for targets having closer neighbors. (d) The SOA strengthens for galaxies in low-density regions, and the signal is dominated by central--satellite pairs in low-mass halos. (e) There is an explicit dependence of the SOA on the duration of interaction with its current neighbor. Taken together, we propose that the SOA witnessed at $z = 0$ has been developed mainly by interactions with a neighbor for an extended period of time, rather than tidal torque from the ambient large-scale structure.
\end{abstract}

%% Keywords should appear after the \end{abstract} command. 
%% See the online documentation for the full list of available subject
%% keywords and the rules for their use.

%% From the front matter, we move on to the body of the paper.
%% Sections are demarcated by \section and \subsection, respectively.
%% Observe the use of the LaTeX \label
%% command after the \subsection to give a symbolic KEY to the
%% subsection for cross-referencing in a \ref command.
%% You can use LaTeX's \ref and \label commands to keep track of
%% cross-references to sections, equations, tables, and figures.
%% That way, if you change the order of any elements, LaTeX will
%% automatically renumber them.
%%
%% We recommend that authors also use the natbib \citep
%% and \citet commands to identify citations.  The citations are
%% tied to the reference list via symbolic KEYs. The KEY corresponds
%% to the KEY in the \bibitem in the reference list below. 

%\NewPageAfterKeywords

\section{Introduction} \label{sec:intro}

The anisotropy of galaxy orientations has been a matter of interest for many decades. Early researchers had already suspected that galaxies have a preferred orientation in the sky \citep{1922MNRAS..82..510R, 1938MNRAS..98..218B, 1939MNRAS..99..534B, 1955AJ.....60..415W}. With the advent of sky surveys, more evidence was discovered to support such an idea. \citet{1964MNRAS.127..517B} reported a systematic alignment of position angles of spiral galaxies in a cluster environment. \citet{1975AJ.....80..477H} provided proof that galaxies in the Coma cluster are preferentially aligned toward the cluster center. Until now, many subsequent observations revealed that there are multiple types of galaxy alignments on a variety of scales \citep[see a review by][and references therein]{2015SSRv..193....1J}. For instance, previous studies found that the spatial distribution of satellites in a galaxy group is elongated along both the major axis of the central galaxy \citep[e.g.,][]{1968PASP...80..252S, 1980MNRAS.191..325C, 1982A&A...107..338B, 2005ApJ...628L.101B, 2006MNRAS.369.1293Y, 2007ApJ...662L..71F} and the direction of the neighboring groups \citep[e.g.,][]{1982A&A...107..338B, 1989ApJ...344..535W, 1994ApJS...95..401P, 2009ApJ...703..951W, 2011MNRAS.414.2029P}. The orientation (or spin) of galaxies appears to be aligned in a specific direction within a group \citep[e.g.,][]{1976ApJ...209...22T, 2005ApJ...627L..21P, 2006ApJ...644L..25A, 2007ApJ...662L..71F, 2018MNRAS.474.4772H} and even within the Local Supercluster \citep[e.g.,][]{1982MNRAS.198..605M, 1986MNRAS.222..525F, 1992PASJ...44..493K, 1993MNRAS.265..874G, 2006Ap&SS.302...43H}.

It is now broadly accepted that an anisotropy of the universe naturally arises within the framework of the standard cosmology. Small density fluctuations in the primordial universe evolve and collapse under gravity into dark matter (DM) halos, which grow hierarchically via repeated mergers and accretions \citep[e.g.,][]{1978MNRAS.183..341W}. The hierarchical growth of halos takes place along preferential directions. Massive halos are fed with material via anisotropic flow from voids to sheets, to filaments, and then to nodes, constituting the large-scale structure (LSS) commonly referred to as the \textit{cosmic web} \citep[e.g.,][]{2014MNRAS.441.2923C}. Furthermore, a tidal field generated by the surrounding LSS acts on the DM halos and galaxies to stretch their shape and induce their spin (\citealt{1969ApJ...155..393P, 1970Ap......6..320D, 1984ApJ...286...38W}; see also \citealt{2009IJMPD..18..173S} for a review). Thus the observed alignments between shape, kinematics, and distribution of galaxies are, so to speak, inevitable outcomes stemming from the geometry of the LSS \citep[e.g.,][]{2001MNRAS.320L...7C, 2001ApJ...559..552C, 2001ApJ...555..106L, 2002MNRAS.332..339P, 2015MNRAS.452.3369C, 2019ApJ...872...37L}.

The alignment between the halo and the LSS is one of the most intensively studied subjects. Theoretical studies based on numerical simulations arrived at a conclusion that there is a mass-dependent alignment between the halo spin and the LSS, suggesting that the spin vector of the less (more) massive halos prefers to be parallel (perpendicular) to the filamentary structure \citep{2007ApJ...655L...5A, 2007MNRAS.381...41H, 2012MNRAS.427.3320C, 2013ApJ...762...72T, 2014MNRAS.440L..46A, 2014MNRAS.443.1090F, 2018MNRAS.481..414G, 2020arXiv200710365G}, which is linked to the primary direction of accretion onto halos \citep[e.g.,][]{2014MNRAS.443.1274L, 2015ApJ...813....6K, 2015MNRAS.446.2744L, 2017MNRAS.468L.123W, 2018MNRAS.473.1562W}. Similarly, the spin axes of halos in sheet environments appear to be parallel to the sheet plane \citep{2007MNRAS.381...41H, 2014MNRAS.440L..46A, 2014MNRAS.443.1090F, 2017MNRAS.468L.123W, 2020arXiv200710365G}. Such theoretical expectations turned out to be in agreement with observations \citep{2013ApJ...775L..42T, 2013MNRAS.428.1827T, 2016MNRAS.457..695P, 2017A&A...599A..31H, 2020MNRAS.492..153B, 2020MNRAS.491.2864W}. Recent progress in hydrodynamic simulations further helps bridge the gap between DM halos and observable galaxies \citep{2014MNRAS.444.1453D, 2014MNRAS.445L..46W, 2015MNRAS.448.3391C, 2018MNRAS.481.4753C, 2018ApJ...866..138W, 2019MNRAS.487.1607G, 2019ApJ...876...52K, 2020MNRAS.493..362K}.

In addition to studying individuals, galaxy pairs are more interesting targets since galaxies in pairs evolve under the influence of the neighboring galaxy as well as that of the surrounding LSS. The halo--LSS alignment implies that the spin of each galaxy and the orbital motion of the system are correlated with the geometry of the LSS, which is not truly random. On top of this, encounters with neighboring galaxies modify the spin and orbit of the galaxies \citep[see, e.g.,][]{2017MNRAS.465.2643C, 2018ApJ...856..114C, 2018MNRAS.477.1567L}. Previous studies have found several types of alignments associated with galaxy pairs in which both the neighbor and the LSS play a role. First, the spins of the two galaxies in a pair tend to be preferentially aligned in the same direction \citep{1993AJ....105..473F, 2004MNRAS.353.1197P, 2009MNRAS.392.1225S, 2010MNRAS.402.1807C, 2014MNRAS.438.1784M, 2018ApJ...858...51K}, although some controversy still remains \citep{1993A&A...272..389O, 2012ApJ...756..135B, 2012ApJ...751..153L}. Such spin--spin alignment of galaxy pairs is generally interpreted as a result of the combined effect of the underlying LSS and the mutual interaction between paired galaxies \citep[e.g.,][]{1979MNRAS.187..287S, 1984ApJ...284..471H}. Second, the line connecting two neighboring galaxies is preferentially along the filament \citep{2015A&A...576L...5T, 2015MNRAS.450.2727T, 2018A&A...619A..24M}, which witnesses that their orbital planes are subject to the large-scale velocity field. Third, some authors have focused on satellite galaxies around a galaxy pair. The current evidence suggests that satellites around pairs are more populated in the region between the pairs than the opposing sides, probably owing to satellites' accretion along filaments connecting the two galaxies in a pair \citep{2016ApJ...830..121L, 2017ApJ...850..132P, 2019MNRAS.488.3100G}. Interestingly enough, \citet{2019MNRAS.484.4325W} recently found that the shape of satellites in the in-between (opposing) region is tangentially (radially) aligned with the direction of the nearby primary galaxy, which is indicative of the tidal effect by the primary pairs.

Recently, a new type of alignment in galaxy pairs, which we refer to as the spin--orbit alignment (SOA), has been introduced. The term `SOA' denotes a spatial alignment between the spin vector and the orbital angular momentum vector of galaxies within a galaxy pair. There have been a few studies providing positive evidence for the SOA, yet the origin remains uncertain. On the observational side, \citet{2019ApJ...872...78L} discovered the dynamical coherence between the spin of target galaxies and the line of sight motion of their neighbors within a distance of 1\,Mpc, based on the data of the \mbox{CALIFA} IFU survey. \citet{2020ApJ...893..154L} found that the optical colors of these coherently moving pairs are more likely to resemble each other compared to the case of anti-coherent pairs, which suggests galaxy interactions between neighbors as a possible origin of the dynamical coherence. However, \citet{2019ApJ...884..104L} showed that the dynamical coherence extends out to several-Mpc distances from the target, which, on the contrary, implies a connection with the LSS. Using a cosmological \textit{N}-body simulation, \citet{2017MNRAS.466.4875L} investigated various kinds of alignments within interacting halos and stated that spin vectors of neighboring pairs are aligned with their orbital angular momentum vectors. \citet{2021ApJ...submit..A} analyzed DM halo pairs with overlapping virial radii and detected a signal of the SOA, which is stronger for gravitationally bound (i.e., merging) pairs. Both studies confirmed the SOA of halo pairs based on DM-only simulations, but so far there has been no attempt to use numerical simulations to examine the SOA on the scale of a galaxy pair.
% \citet{2019ApJ...884..104L} argued that the coherence may have multiple origins, both inherited and induced, depending on the scales of interest.
% \citet{2019ApJ...872...78L} showed that the coherence is particularly strong for faint, blue, and kinematically misaligned galaxies. 

The SOA can be understood in the context of the effects of both LSS and neighbors. As already addressed, previous studies have found the alignment between the galaxy spin and the LSS \citep[e.g.,][]{2013ApJ...775L..42T, 2013MNRAS.428.1827T, 2014MNRAS.444.1453D}, which is attributed to the large-scale tidal field \citep[e.g.,][]{2009IJMPD..18..173S} and the preferential accretion of material \citep[e.g.,][]{2014MNRAS.443.1274L, 2015ApJ...813....6K}. With the proximity of two galaxies in a pair, a common local tidal field determines their angular momenta in the early universe. The anisotropic accretion along the cosmic web naturally leads to the SOA because the galaxy spin is influenced by the orbital angular momentum of the accreted matter. On the other hand, a well-known example of the effect of interactions on the galaxy orientation is given by satellites in massive groups. Due to the tidal effect of the central halo, the major axis of the satellite is eventually aligned to the direction of the halo center \citep{1997ApJ...487..489U, 2008ApJ...672..825P, 2015MNRAS.451.2536R}, which is called the radial alignment of satellites and has been investigated in many studies through both observations \citep{1976ApJ...209...22T, 2005ApJ...627L..21P, 2006ApJ...644L..25A, 2007ApJ...662L..71F, 2018MNRAS.474.4772H, 2019ApJ...883...56R} and theoretical models \citep[][]{1994MNRAS.270..390C, 1997ApJ...487..489U, 2007ApJ...671.1135K, 2008ApJ...672..825P, 2015MNRAS.451.2536R, 2015MNRAS.448.3522T, 2020MNRAS.495.3002K}. Similarly, \citet{2018A&A...613A...4W} showed, using a cosmological simulation, that the alignment between the spin of a central galaxy and the orbits of its satellites is developed in the inner region of the halo by gravitational torques from the central, while the dynamics of satellites in the outer region is more governed by the geometry of the LSS.

In this series of papers, we have highlighted the impact of interacting neighboring galaxies (or halos) on galaxy evolution by means of both observations and theoretical models. \citet[][Paper \citetalias{2019ApJ...882...14M}]{2019ApJ...882...14M} investigated the impact of galaxy interactions on star formation based on the Sloan Digital Sky Survey and showed that the star formation activity is enhanced or reduced depending on neighbors' hydrodynamic properties. \citet[][Paper \citetalias{2019ApJ...887...59A}]{2019ApJ...887...59A}, using a set of cosmological \textit{N}-body simulations, found that the dominance of flyby interactions over mergers increases at lower redshifts, for less massive halos, and in denser environments. In the present paper, we explore the SOA of galaxy pairs using a state-of-the-art cosmological hydrodynamic simulation from the IllustrisTNG project \citep{2018MNRAS.480.5113M, 2018MNRAS.477.1206N, 2018MNRAS.475..624N, 2018MNRAS.475..648P, 2018MNRAS.475..676S}. This is the first study, to our knowledge, to utilize a cosmological hydrodynamic simulation to investigate the SOA of galaxy pairs. Our main goal is to elucidate whether the SOA on a galaxy-pair scale is present in the IllustrisTNG and, if so, how it depends on the intrinsic and environmental properties of galaxies. Taking full advantage of the baryonic physics model, we here particularly focus on the baryonic component (i.e., stars and gas) of galaxies and discuss the possible formation mechanisms of the SOA. 

This paper is structured as follows. Section~\ref{sec:methods} describes the selection of galaxy pairs from the IllustrisTNG project and the method to measure the spin--orbit angle. Section~\ref{sec:results} presents the main results, including the analysis of the SOA and its dependence on galaxy properties. We discuss the physical implication of our results in Section~\ref{sec:discussion} and conclude our analysis in Section~\ref{sec:conclusion}. 

\section{Methods} \label{sec:methods}

\subsection{The IllustrisTNG Simulation} \label{subsec:simul}

For the present study, we use the public data release of the IllustrisTNG project \citep{2019ComAC...6....2N}. The IllustrisTNG project is a suite of cosmological magnetohydrodynamical simulations performed with the moving-mesh code \texttt{AREPO} \citep{2010MNRAS.401..791S}. The simulation assumes a $\Lambda$CDM cosmology with $\Omega_m\,=\,0.3089$, $\Omega_\Lambda\,=\,0.6911$, $\Omega_b\,=\,0.0486$, $h\,=\,0.6774$, $\sigma_8\,=\,0.8159$, and $n_s\,=\,0.9667$ \citep{2016A&A...594A..13P}. A set of physical models describing galaxy formation and evolution is employed in the simulation suite, including primordial and metal-line radiative cooling, star formation, chemical enrichment, stellar feedback-driven outflow, supermassive black hole growth, and active galactic nucleus feedback \citep[for details, see][]{2017MNRAS.465.3291W, 2018MNRAS.473.4077P}. The model reasonably well reproduces the fundamental properties and scaling relations of observed galaxy populations \citep{2018MNRAS.480.5113M, 2018MNRAS.477.1206N, 2018MNRAS.475..624N, 2018MNRAS.475..648P, 2018MNRAS.475..676S}.

We use the TNG100 run \citep[see for details][]{2019ComAC...6....2N}, which has a box with a side length of 75 $h^{-1}$Mpc. The data release contains three realizations of the same volume at different resolution levels (TNG100-1, -2, and -3) and their dark matter-only counterparts (-Dark). The main results presented here are based on the highest-resolution run (TNG100-1), which includes 1820$^3$ DM particles and an equal number of initial gas cells, and the corresponding mass resolution is 7.5 $\times$ 10$^6$ $M_\odot$ and 1.4 $\times$ 10$^6$ $M_\odot$ for dark matter and baryonic components, respectively. The gravitational softening length for collisionless particles is 1.0 $h^{-1}$kpc in comoving units and is further limited to 0.5 $h^{-1}$kpc in physical units below $z = 1$. The softening of gas cells is adaptively scaled to be proportional to the cell radius. The simulation starts from $z = 127$ and outputs 100 snapshots from $z = 20$ to $z = 0$ with a maximum time spacing of $\sim$\,200 Myr. 

The data provided in each snapshot includes the halo catalog and the merger tree. Dark matter halos are identified by the friends-of-friends (FoF) algorithm with a linking length of 0.2 times the mean interparticle separation. Subhalos, locally overdense and self-bound substructures within the FoF halos, are detected by using the \texttt{SUBFIND} code \citep{2001MNRAS.328..726S, 2009MNRAS.399..497D}. Each subhalo with nonzero stellar mass corresponds to a single galaxy, making each FoF halo equivalent to a group or a cluster of galaxies. We refer to the most massive subhalo in each FoF halo as a central and to other subhalos as satellites. The merger tree enables a subhalo at any epoch to be connected to its progenitors or descendants at different snapshots. The \texttt{SubLink} code \citep{2015MNRAS.449...49R} generates the merger tree used in this study.

\subsection{Sample Selection} \label{subsec:sample}

We start by defining a sample of paired galaxies from the TNG100. To avoid selecting spurious objects, we only consider subhalos with the stellar mass $M_*$ $>$ 10$^8$ $h^{-1}M_\odot$ at $z = 0$, which typically contains at least 100 star particles. The sample is also restricted to subhalos that are identified as of a cosmological origin in the \texttt{SubhaloFlag} field of the subhalo catalog \citep{2019ComAC...6....2N}, excluding non-cosmological objects such as overdense clumps embedded within galaxies. By definition, a subhalo is flagged as a cosmological one if it is formed either (\romannumeral 1) as a central, (\romannumeral 2) outside the virial radius of its host halo, or (\romannumeral 3) with the dark matter fraction higher than 0.8. We consider these luminous, cosmological subhalos as galaxies throughout the paper. 

The nearest neighboring galaxy is identified for all target galaxies with $M_*$ $>$ 10$^9$ $h^{-1}M_\odot$ at $z = 0$. The nearest neighbor should have a stellar mass larger than at least one-tenth of the target's mass to ensure a strong influence exerted on the target galaxy. Once the identification is made, we limit our analysis to pairs of galaxies with the stellar mass ratio between the target and neighbor, $|\mu_*| = |$log($M_{*, \mathrm{nei}}/M_{*, \mathrm{target}}$)$| < 1.0$, because our interest is in paired galaxies of comparable mass, not satellites around a giant galaxy. When a target galaxy and its nearest neighbor are both members of the same FoF halo, the target is regarded as paired with the neighbor and used in our analysis. In total, the sample contains 7607 target galaxies at $z = 0$. We note that not all the targets are in isolated binary systems; only 3494 galaxies out of the 7607 are mutually paired with each other. Note also that we do not consider whether the nearest neighbor is a central or a satellite during the sample selection process, and thus our sample includes both central--satellite and satellite--satellite pairs; 2681 and 4926 galaxies belong to each group, respectively.

\subsection{Remarks on Paired Galaxies} \label{subsec:pairedgalaxies}

Our sample selection is on the basis of the subhalo properties identified by the halo finder, but the ability of halo finders is known to be incomplete for interacting systems \citep[e.g.,][]{2011MNRAS.410.2617M, 2013MNRAS.435.1618K, 2015MNRAS.454.3020B}. Specifically, the \texttt{SUBFIND} algorithm, which is used in this study, assigns particles unbound to any substructure in an FoF halo to the background halo \citep[see][]{2001MNRAS.328..726S}. This causes loosely bound particles in the outskirt of the smaller of the two interacting galaxies (i.e., identified as a satellite) are assigned to the larger (i.e., a central). Hence, the membership of particles to subhalos is severely affected by the presence of a companion, and the spurious mass loss occurring in satellites increases the mass ratio of galaxy pairs as the two galaxies approach each other \citep{2015MNRAS.454.3020B, 2015MNRAS.449...49R, 2020MNRAS.494.4969P}. Fortunately, as shown in \citet{2015MNRAS.454.3020B}, the position of a subhalo, defined as the location of the particle with the lowest gravitational potential, is relatively robust as long as the halo finder detects the subhalo. The stellar mass, however, is very vulnerable to numerical mass loss. 

\begin{figure*}[tb!]
\epsscale{1.17}
\plotone{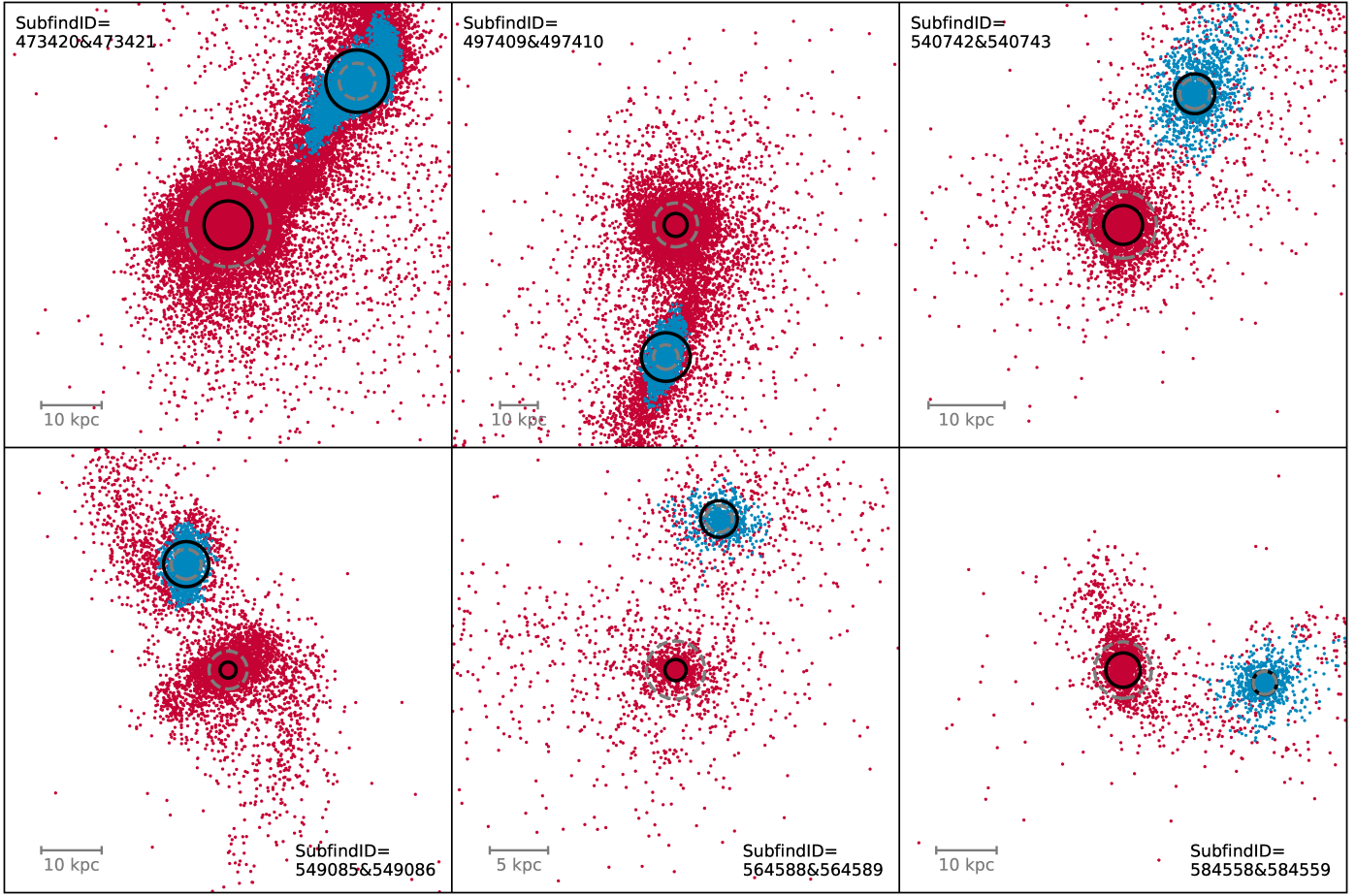}
\caption{Examples of paired galaxies in TNG100. The selected six pairs are equal-mass (1/3 $<$ $\mu_*$ $<$ 3) central--satellite pairs where the central shows the largest difference in stellar mass from the original catalog. Star particles assigned in centrals and satellites are colored red and blue, respectively. In each panel, the original stellar half-mass radius is shown as a gray dashed circle, and the updated radius as a black circle. \label{fig:pair_example}}
\end{figure*}

We update the stellar mass, $M_*$, of each paired galaxy as the value at the snapshot where the less massive of the two galaxies in a pair reaches its maximum stellar mass ($t_{\mathrm{max}}$), as suggested in \citet{2015MNRAS.449...49R}. We then calculate the stellar half-mass radius, $r_\mathrm{h}$, which is defined as the radius of a sphere containing half of the stellar mass. The calculation of $r_\mathrm{h}$ is based on the stellar mass at $t_{\mathrm{max}}$ and the distribution of star particles at $z = 0$ without using the particle membership information obtained by the halo finder. Here, unlike the original definition of $t_{\mathrm{max}}$, we define the time when a galaxy is paired with its current neighbor ($t_{\mathrm{nei}}$) and limit the search of $t_{\mathrm{max}}$ to snapshots (\romannumeral 1) after $t_{\mathrm{nei}}$ and (\romannumeral 2) within the past 2 Gyr. These conditions are applied to prevent too early $t_{\mathrm{max}}$ for galaxies that gradually lose mass regardless of interactions with their current neighbors.\footnotenew{For the same reason, \citet{2020MNRAS.494.4969P} defined $t_{\mathrm{max}}$ as the time when a galaxy reaches its maximum stellar mass within the past 0.5 Gyr. We instead use $t_{\mathrm{nei}}$ as a constraint to take into account the difference in interaction time for each pair, but $\sim$\,90\,\% of targets have $t_{\mathrm{max}}$ within the past 0.5 Gyr. Additionally, we impose an upper limit of 2 Gyr in lookback time to minimize the effect of physical stripping, which affects only $\sim$\,2\,\% (167 targets) of the sample.} In practice, we determine $t_{\mathrm{nei}}$ by finding the earliest snapshot where the nearest neighbor at $z = 0$ remains the same without any change, paying special attention to the following two cases. First, for galaxies with broken merger trees, which are mostly ones deprived of DM, we repair the trees by filling in the missing progenitors with subhalos having the largest number of star particles in common. Second, some galaxies always have the same nearest neighbor since their stellar masses reach 10$^9$ $h^{-1}M_\odot$, below which we cannot correctly identify the neighbor at a given mass resolution. For these galaxies, the earliest snapshot where the neighbor is identified (i.e., the time when the stellar mass exceeds 10$^9$ $h^{-1}M_\odot$) is used as (the lower limit of) $t_{\mathrm{nei}}$.

\begin{figure*}[tb!]
\epsscale{1.18}
\plotone{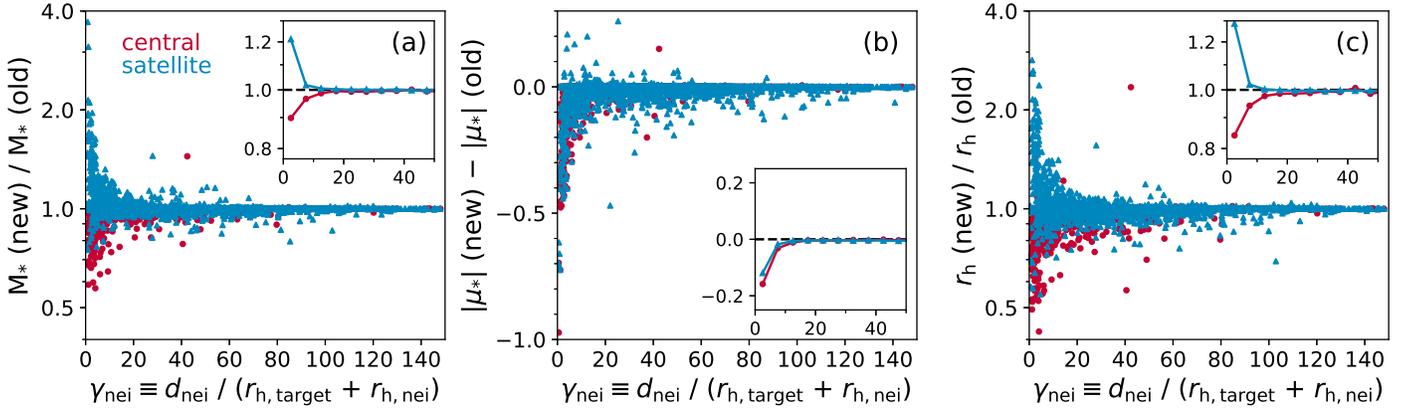}
\caption{Comparison between the updated and the catalog values of (a) the stellar mass, (b) the stellar mass ratio, and (c) the stellar half-mass radius with respect to the separation to the nearest neighbor for centrals (red circle) and satellites (blue triangle) in the paired galaxy sample. The insets show the mean changes in the subhalo properties at a small separation. \label{fig:newassign}}
\end{figure*}

Figure~\ref{fig:pair_example} demonstrates several examples of close galaxy pairs in TNG100. The six pairs are selected among central--satellite pairs that have a large difference between the stellar mass at $t_{\mathrm{max}}$ and at $z = 0$. Star particles in centrals and satellites are shown in red and blue, respectively, according to the membership information determined by the halo finder at $z = 0$. It is clear that, for these pairs, the halo finder assigns stars in the outer part of the satellites as members of the centrals. Consequently, this makes the updated $r_\mathrm{h}$ of centrals smaller and that of satellites larger than the catalog values, as seen by comparing the solid and dashed circles. We compare the updated mass and size with those stored in the halo catalog in Figure~\ref{fig:newassign}. Panel (a) shows the ratio of the new to the original stellar masses with respect to the separation to the nearest neighbor, $\gamma_\mathrm{nei}$, defined as
\begin{equation}
\gamma_\mathrm{nei} = \frac{d_\mathrm{nei}}{r_\mathrm{h,target} + r_\mathrm{h,nei}},
\end{equation}
where $d_\mathrm{nei}$ is the 3D distance to the nearest neighbor, and $r_\mathrm{h,target}$ and $r_\mathrm{h,nei}$ are the stellar half-mass radii of the target and the nearest neighbor, respectively. For centrals (red), the updated stellar mass decreases below the catalog value as the neighbor approaches. On the contrary, the updated stellar mass of satellites (blue) increases as the separation decreases. The discrepancy in the stellar mass starts to appear at $\gamma_\mathrm{nei}\sim 20$ but becomes significant at $\gamma_\mathrm{nei} < 10$, for both centrals and satellites (see the inset). The decrease in the centrals' mass and the increase in the satellites' mass result in the smaller mass ratio, as presented in panel (b). Similarly, the change in the total stellar mass is followed by the change in the stellar half-mass radius in panel (c). We stress again that, although there may be some real effect of interactions, the mass transfer between the central and satellite is primarily numerical and not physical in most cases, as discussed previously.

\subsection{Spin--Orbit Angle} \label{subsec:spin-orbit}

To quantify the strength of the SOA, we first calculate the spin vector of a target and the orbital angular momentum vector of its neighbor. The spin vector, $\textit{\textbf{S}}$, denotes an angular momentum vector within the stellar half-mass radius, that is,
\begin{equation}
\textit{\textbf{S}} = \sum_i\ m_i\,(\textbf{\textit{r}}_i \times \textbf{\textit{v}}_i)\quad\mathrm{only\ if}\ r_i < r_\mathrm{h},
\end{equation}
where $m_i$ is the mass of the $i$th star particle, and $\textbf{\textit{r}}_i$ and $\textbf{\textit{v}}_i$ are the position and velocity with respect to the galaxy center, respectively. As shown in Figure~\ref{fig:pair_example}, many paired galaxies are more or less overlapped with each other, making it difficult to separate the two galaxies. For this reason, we choose to use only the spin at the central region within $r_\mathrm{h}$ determined in Section~\ref{subsec:pairedgalaxies}. We measure the spin vector ($\textit{\textbf{S}}$) separately for each particle type (gas, star, and DM) with the same size of aperture, considering the misalignment between them \citep[e.g.,][]{2017MNRAS.472.1163C, 2020MNRAS.492.1869D, 2020ApJ...894..106K}. The orbital angular momentum vector, $\textit{\textbf{L}}$, is calculated by taking the cross product of the position vector ($\textbf{\textit{d}}_\mathrm{nei}$) and the velocity vector ($\textbf{\textit{v}}_\mathrm{nei}$) of the nearest neighbor in the reference frame of the target, namely
\begin{equation}
\textit{\textbf{L}} = \textbf{\textit{d}}_\mathrm{nei} \times \textbf{\textit{v}}_\mathrm{nei}.
\end{equation}
In all the calculations, the position of each galaxy center is defined as the location of the particle at the lowest potential, and the velocity is the mass-weighted mean velocity of all particles within $r_\mathrm{h}$.

\begin{figure}[tb!]
\epsscale{1.17}
\plotone{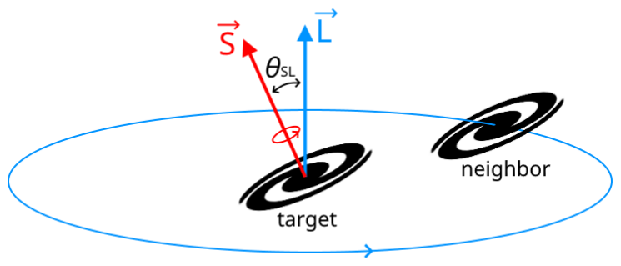}
\vspace{-10mm}
\caption{Definition of the spin--orbit angle \thetasl{}, which is the angle between the spin vector of the target $\textit{\textbf{S}}$ and the orbital angular momentum vector of the neighbor $\textit{\textbf{L}}$. \label{fig:soorient}}
\end{figure}

The spatial resolution of TNG100 ($\sim$0.7 kpc for star and DM particles at $z = 0$) is not satisfactory to resolve the inner structure of galaxies. Nevertheless, about 98\,\% of galaxies in our sample have $r_\mathrm{h}$ larger than the softening length for star particles, and about 60\,\% of the galaxies have $r_\mathrm{h}$ exceeding 2.8 times the softening length where the force is exactly Newtonian. We need only the spin direction within $r_\mathrm{h}$ for our analysis, and therefore we do not expect the resolution changes the overall results. To obtain a more reliable measurement of the spin vector, we require a minimum of 100 elements for each particle type within $r_\mathrm{h}$. Note that we also exclude 173 severely overlapped pairs with $\gamma_\mathrm{nei} < 3$ from the sample to prevent contamination from particles belonging to the neighbor. The criteria leave 4226, 7287, and 7434 galaxies with an available spin vector for gas, DM, and stars, respectively.

\begin{figure*}[tb!]
\epsscale{1.18}
\plotone{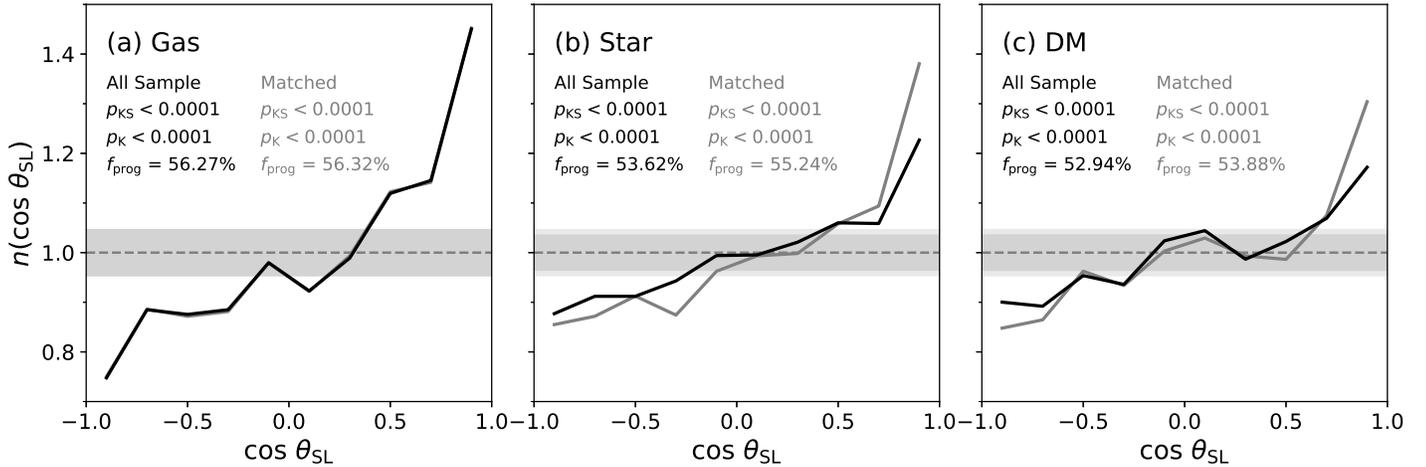}
\caption{PDF of the spin--orbit angle of paired galaxies in TNG100 for (a) gas, (b) stars, and (c) DM. Black lines show the PDFs for galaxies that have at least 100 corresponding cells or particles within $r_\mathrm{h}$. The shaded region shows the standard deviation estimated from 1000 randomly generated isotropic samples with the same size as the original sample. Gray lines show the PDFs for galaxies with at least 100 cells or particles available for all the three components of galaxies within $r_\mathrm{h}$. The two lines are almost identical in panel (a). The $p$-values of the KS test ($p_\mathrm{KS}$) and the Kuiper's test ($p_\mathrm{K}$) and the prograde fraction ($f_\mathrm{prog}$) are given on each panel. All $p$-values less than 10$^{-4}$ are denoted as $<$ 0.0001. \label{fig:soa}}
\end{figure*}

We define the spin--orbit angle \thetasl{} as the angle between the spin vector ($\textit{\textbf{S}}$) and the orbital angular momentum vector ($\textit{\textbf{L}}$) in a pair (see Figure~\ref{fig:soorient}). Note that the angle, \thetasl{}, is measured from $\textit{\textbf{S}}$ to $\textit{\textbf{L}}$ in the counterclockwise direction when viewed from the positive z-side, within the range from 0$^{\circ}$ (aligned) to $\pm$180$^{\circ}$ (anti-aligned). We refer to the case where \costheta{} $>$ 0 (i.e., $|$\thetasl{}$|$ $<$ 90$^{\circ}$) as `prograde' and the case where \costheta{} $<$ 0 (i.e., $|$\thetasl{}$|$ $>$ 90$^{\circ}$) as `retrograde,' respectively. The probability distribution function (PDF) of the spin--orbit angle is expressed as
\begin{equation}
n(\mathrm{cos}\,\theta_\mathrm{SL}) = \frac{N(\mathrm{cos}\,\theta_\mathrm{SL})}{\langle N_{\mathrm{rand}}(\mathrm{cos}\,\theta_\mathrm{SL}) \rangle},
\end{equation}
where $N$(\costheta{}) is the number of pairs in each bin, and $\langle N_{\mathrm{rand}}$(\costheta{})$\rangle$ is the expected number of pairs from the random isotropic sample with the same size. Therefore, $n$(\costheta{}) is close to unity in a uniform distribution, and, if paired galaxies have a preference for a spin--orbit angle of \thetasl{}, $n$(\costheta{}) would be greater than 1. We compute the standard deviation of PDFs, $\sigma_{\mathrm{rand}}$(\costheta{}), over 1000 random isotropic samples and, following \citet{2006MNRAS.369.1293Y}, use $\sigma_{\mathrm{rand}}$(\costheta{})$/\langle N_{\mathrm{rand}}$(\costheta{})$\rangle$ to evaluate the significance of the alignment. 

\section{Results} \label{sec:results}

\subsection{Spin--Orbit Alignment} \label{subsec:soatng}

Figure~\ref{fig:soa} shows the PDF of the spin--orbit angle, $n$(\costheta{}), for paired galaxies in TNG100. Black lines in panels (a), (b), and (c) represent the results obtained for gas, stars, and DM, respectively. It is obvious that the paired galaxies prefer prograde orientations to retrograde ones; $n$(\costheta{}) is greater than 1 for \costheta{} $\gtrsim$ 0 and less than 1 for \costheta{} $\lesssim$ 0. The preference of prograde orientation is common in all panels. The maximum significance compared to the random isotropic sample is 9.8$\sigma$, 6.5$\sigma$, and 4.9$\sigma$ at \costheta{} $\sim$ 1 for gas, stars, and DM, respectively. We conduct Kolmogorov--Smirnov (KS) tests using the \texttt{SciPy} module \citep{2020NMeth..17...261V} to quantify the statistical difference of the PDF of the spin--orbit angle from the uniform distribution. The $p$-values of the hypothesis that the samples in panels (a)--(c) are drawn from the random orientation are all less than 10$^{-4}$ ($p_\mathrm{KS}$ $<$ 0.0001).\footnotenew{We denote the $p$-values less than 10$^{-4}$ as `$<$ 0.0001' throughout this paper since there is no point in presenting $p$-values to as many decimal places as possible, due to their high sample-to-sample variability \citep[see, e.g.,][]{2011TAS....65..213B, 2014MP.....19.1336L}.} While the widely used KS test strongly suggests that the spin--orbit orientation is not random, the sensitivity of the KS test is not uniform across the entire distribution (i.e., most sensitive near the median), and hence it may not be adequate for directional data (e.g., angles and vectors). Therefore, we additionally test the isotropy using the Kuiper's test \citep[see][]{1965Biome..52..309S, 2004A&A...420..789P} in the \texttt{astropy} module \citep{2018AJ....156..123A}, which is similar to the KS test but rotation-invariant. The probability that the distribution of \thetasl{} is drawn from isotropic $\textit{\textbf{S}}$ and $\textit{\textbf{L}}$ (against a sinusoidal PDF of $n$(\thetasl{}) $\sim$ $|$sin(\thetasl{})$|$) is found to be very low ($p_\mathrm{K}$ $<$ 0.0001) in all three panels. Also, we compute the prograde fraction, $f_\mathrm{prog}$, as a simple metric to quantify the prevalence of prograde orientations, such that
\begin{equation}
f_\mathrm{prog} = \frac{N(\mathrm{cos}\,\theta_\mathrm{SL} > 0)}{N_\mathrm{total}},
\end{equation}
where $N$(\costheta{} $>$ 0) is the number of galaxies on prograde orbits and $N_\mathrm{total}$ is the total number of galaxies, and the expected value of $f_\mathrm{prog}$ for the isotropic orientation is 50\,\%. The prograde alignment for the gas spin vector is the strongest, and the strength decreases in the order of gas, stars, and DM; $f_\mathrm{prog}$ is 56.3 $\pm$ 0.76\,\%, 53.6 $\pm$ 0.58\,\%, and 52.9 $\pm$ 0.58\,\% for each component, respectively.

We note that each panel has a different sample size since we restrict the sample to galaxies that contain at least 100 corresponding cells or particles within $r_\mathrm{h}$. Gas-poor galaxies usually contain a small number of gas cells, making the sample size of panel (a) smaller than the others. We repeat the analysis using galaxies commonly included in all three panels to remove the selection effect and display the result as gray lines in Figure~\ref{fig:soa}. For this `matched' sample, the result does not change significantly. There is a clear signal of prograde alignment, and the strongest alignment is obtained for gas, followed by stars, and the weakest is for DM. The only notable difference is that the prograde fraction for stars and DM become slightly higher than that for the whole sample; $f_\mathrm{prog}$ $=$ 56.3\,\%, 55.2\,\%, and 53.9\,\% with $\sigma$ $=$ $\sqrt{f(1-f) / N}$ $\sim$ 0.77\,\% for gas, stars, and DM in the `matched' sample, respectively. This increase is because, as we will see in the next section, the alignment signal for gas-rich galaxies, which generally reside in sparse environments, is stronger than that for gas-poor galaxies in dense environments.

\begin{figure*}[tb!]
\epsscale{1.18}
\plotone{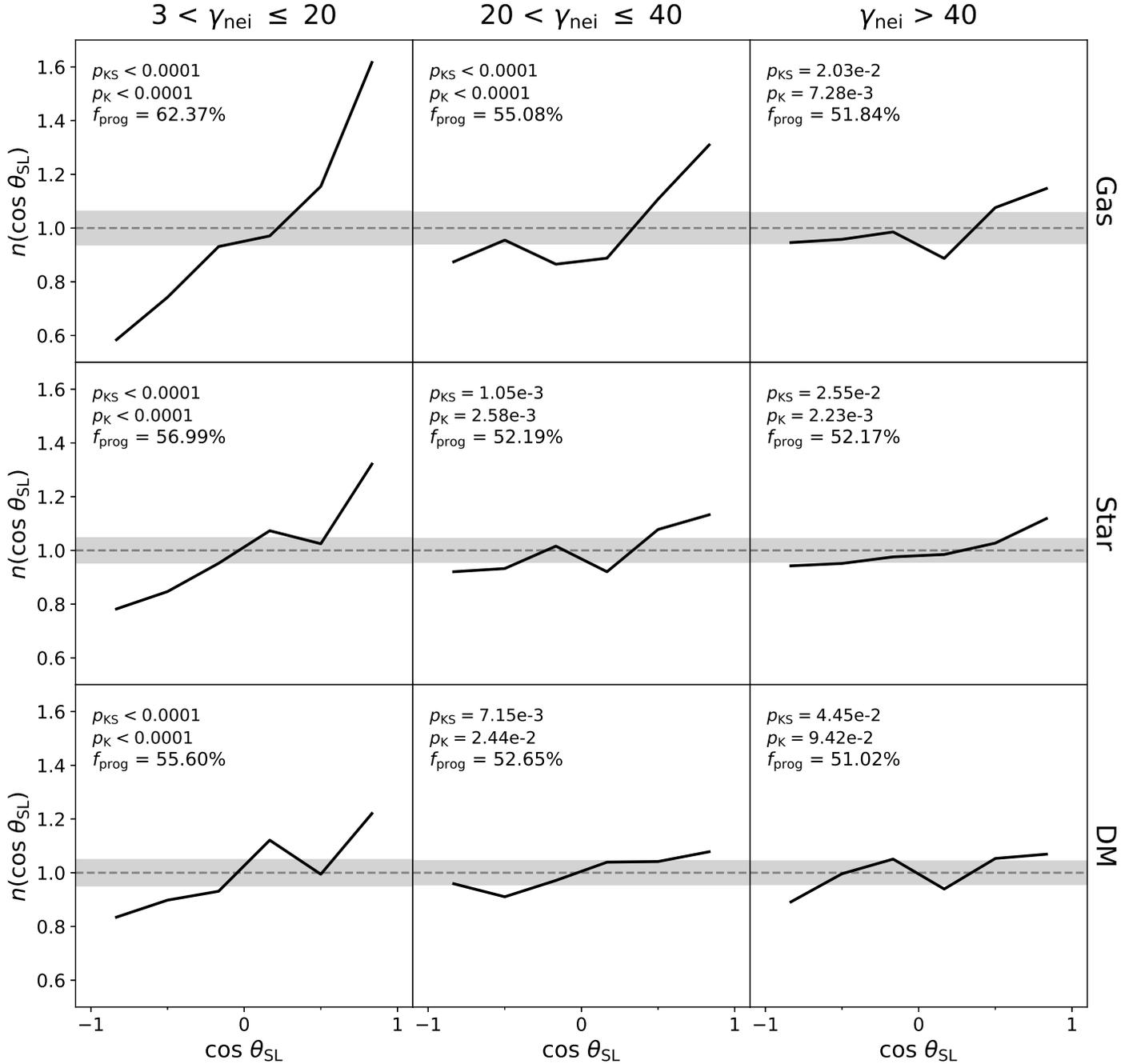}
\caption{PDF of the spin--orbit angle of paired galaxies in subsamples split by separation. From left to right, the separation to the neighbor $\gamma_\mathrm{nei}$ increases. The spin--orbit angle of each galaxy is calculated separately for gas (top), stars (middle), and DM (bottom). The format is the same as in Figure~\ref{fig:soa}. \label{fig:soadist}}
\end{figure*}

Doing this analysis, we are careful to ensure that the alignment signal is real and not an artifact of the uncertainty in the particle membership. If some particles belonging to the neighbor were slid into the target, their orbital angular momenta would be added to the target's spin, which would result in spurious alignment. However, we remind the reader that the spin vector is measured within $r_\mathrm{h}$, and the galaxy pairs should have $\gamma_\mathrm{nei}> 3$---that is, the distance between the two galaxies should be greater than six times of the mean $r_\mathrm{h}$ of the two, while, on average, more than 90\,\% of their stellar mass is enclosed within $5 \times r_\mathrm{h}$ \citep[see also][]{2018MNRAS.475..648P}. Furthermore, we expect that the problem should be most prominent for DM, which is more spatially extended than the other components, but the alignment for the DM appears to be the weakest in Figure~\ref{fig:soa}. Therefore, we conclude that at least the SOA for stars is reliable, although one must keep in mind that the alignment for more extended components like gas and DM can be artificially exaggerated. 

\begin{figure*}[tb!]
\epsscale{1.18}
\plotone{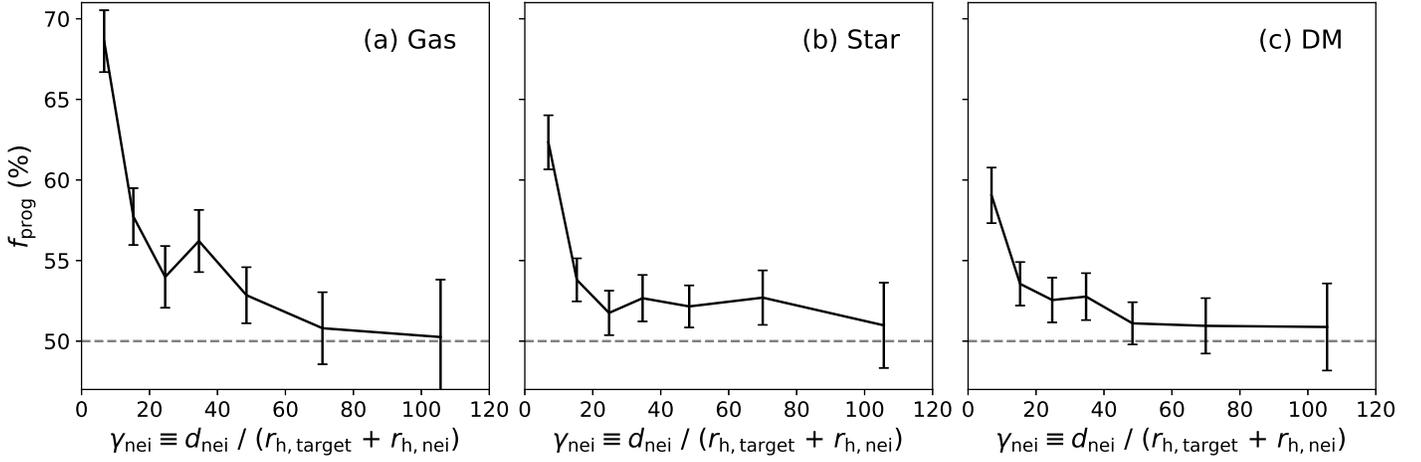}
\caption{Prograde fraction of the spin--orbit angle calculated for (a) gas, (b) stars, and (c) DM as a function of the separation to the nearest neighbor. Error bars represent the standard error of the percentage. \label{fig:fprogdist}}
\end{figure*}

Figure~\ref{fig:soadist} shows the PDF of the spin--orbit angle as a function of the separation to the nearest neighbor ($\gamma_\mathrm{nei}$). The sample is split into four subsamples by separation to the neighbor. The figure clearly shows that the alignment signal becomes weaker as the separation increases. There is a significant excess in the fraction of prograde orientations for paired galaxies with $\gamma_\mathrm{nei} < 20$, and the $p$-values from the KS and the Kuiper's tests for this subsample are extremely small ($p < 0.0001$) for all the components of gas, stars, and DM. The prograde fraction decreases with increasing separation, and the $p$-values also increase. This indicates that a strong signal of the SOA requires the presence of close neighbors. In Figure~\ref{fig:fprogdist}, we plot the prograde fraction as a function of the separation to the nearest neighbor. First of all, the prograde fraction is larger than 50\,\% even at a relatively large separation ($\gamma_\mathrm{nei} \sim 50$), which suggests that paired galaxies generally prefer prograde orientations rather than random ones. As already seen in Figure~\ref{fig:soadist}, the prograde fraction decreases as the separation increases. While the spin direction of galaxies with $\gamma_\mathrm{nei} > 100$ are almost randomly distributed with respect to the orbit, the prograde fraction sharply increases at $\gamma_\mathrm{nei} < 20$ and reaches to 68.6 $\pm$ 1.92\,\%, 62.3 $\pm$ 1.68\,\%, and 59.0 $\pm$ 1.72\,\% at $3 < \gamma_\mathrm{nei} < 10$ (the leftmost bin) for gas, stars, and DM, respectively. 

The SOA that we find here is in line with previous studies \citep{2017MNRAS.466.4875L, 2019ApJ...872...78L}. For instance, \citet{2017MNRAS.466.4875L} discovered the SOA for paired galaxies within the virial radius of the neighboring halo (on a scale of hundreds of kiloparsecs), using an \textit{N}-body simulation. \citet{2019ApJ...872...78L} observationally found the kinematic coherence within a distance of 1\,Mpc. Back in Figures~\ref{fig:soadist} and \ref{fig:fprogdist}, the alignment signal extends out to $\gamma_\mathrm{nei} \sim 100$, which corresponds to $d_\mathrm{nei} \sim 600$\,kpc on average with a large spread, and hence the scale of the alignment seen in this study is roughly comparable to that of previous results. Going a step further, we make use of a cosmological hydrodynamic simulation to study the SOA of galaxy pairs for the first time. The result from the IllustrisTNG shows that a strong prograde alignment is established even if baryonic processes are taken into account.

\subsection{Mass and Environment Dependence} \label{subsec:dependme}

In order to deepen our understanding of the SOA, we explore the factors that influence the strength of the alignment signal. Galaxy mass and environment are key factors that play a role in galaxy evolution, each representing the internal and external aspects of galaxy characteristics. We thus begin by examining the dependence of the SOA on mass and environment. As for the mass, we simply use the stellar mass of the target, $M_*$, defined by the method described in Section~\ref{subsec:pairedgalaxies}. As for the environment, we calculate the number density of galaxies, which is defined as
\begin{equation}
\Sigma_k = \frac{3k}{4 \pi (d_k / \mathrm{Mpc})^3},
\end{equation}
where $d_k$ is the distance to the $k$th nearest neighbor from the target. To calculate the density, we consider only luminous subhalos of the cosmological origin (see Section~\ref{subsec:sample}). We take the geometric mean of $\Sigma_4$ and $\Sigma_5$ as a representative estimate of the local density $\Sigma$. 

\begin{figure*}[tb!]
\epsscale{1.17}
\plotone{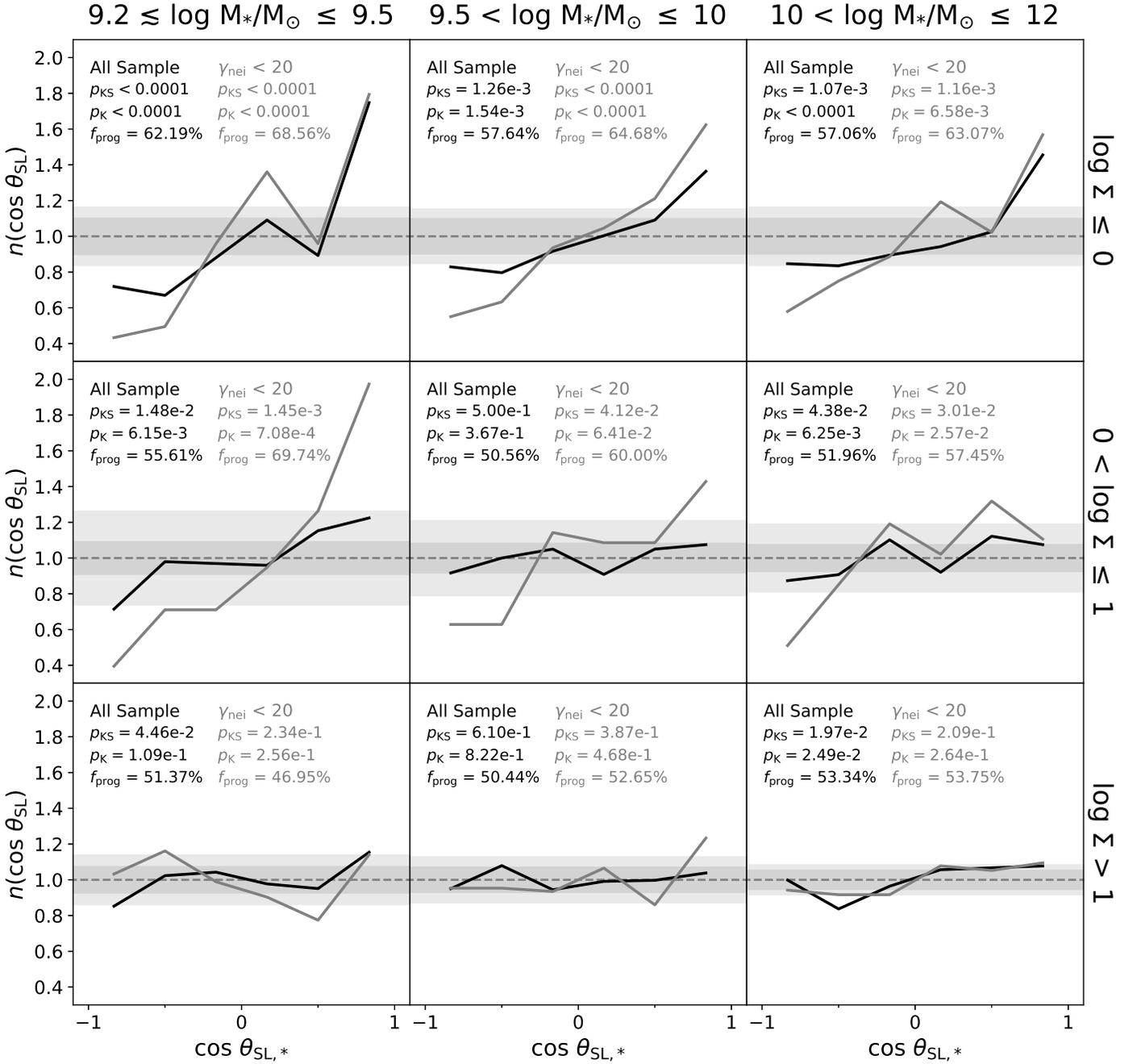}
\caption{PDF of the stellar spin--orbit angle of paired galaxies in subsamples split by the stellar mass and the local density. The stellar mass $M_*$ increases from left to right columns, and the local density $\Sigma$ increases from top to bottom rows. Black lines and dark gray shades are for the whole sample, and gray lines and light gray shades are for paired galaxies with $\gamma_\mathrm{nei}$ $<$ 20. The format is the same as in Figure~\ref{fig:soa}. \label{fig:soame}}
\end{figure*}

Figure~\ref{fig:soame} shows the PDF of the stellar spin--orbit angle for nine subsamples split by the stellar mass and the local density. We present only the result for stars, which is the most robust, but similar trends are also seen for gas and DM. We find that the significance of alignment increases as both the stellar mass and the local density decrease. Low-mass target galaxies in low-density environments (shown in the top left panel) exhibit the strongest alignment signal ($p_\mathrm{KS}$ $<$ 0.0001, $p_\mathrm{K}$ $<$ 0.0001 and $f_\mathrm{prog}$ $=$ 62.2 $\pm$ 2.20\,\%). Of the two parameters, the local density seems to have a greater impact on the alignment than the stellar mass in given ranges. For instance, while low-mass galaxies in dense environments (bottom left) show no clear evidence for the alignment ($f_\mathrm{prog}$ $=$ 51.4 $\pm$ 1.65\,\%), massive galaxies in sparse environments (top right) have a moderate preference for prograde orientations ($f_\mathrm{prog}$ $=$ 57.1 $\pm$ 2.21\,\%). We emphasize that the dependence of the SOA on the mass and environment is not due to the difference of the separation to the neighbor in each subsample. We obtain the same trend when repeating the analysis for a fixed range of the separations ($\gamma_\mathrm{nei} < 20$; shown as gray lines).

\begin{figure*}[tb!]
\epsscale{1.17}
\plotone{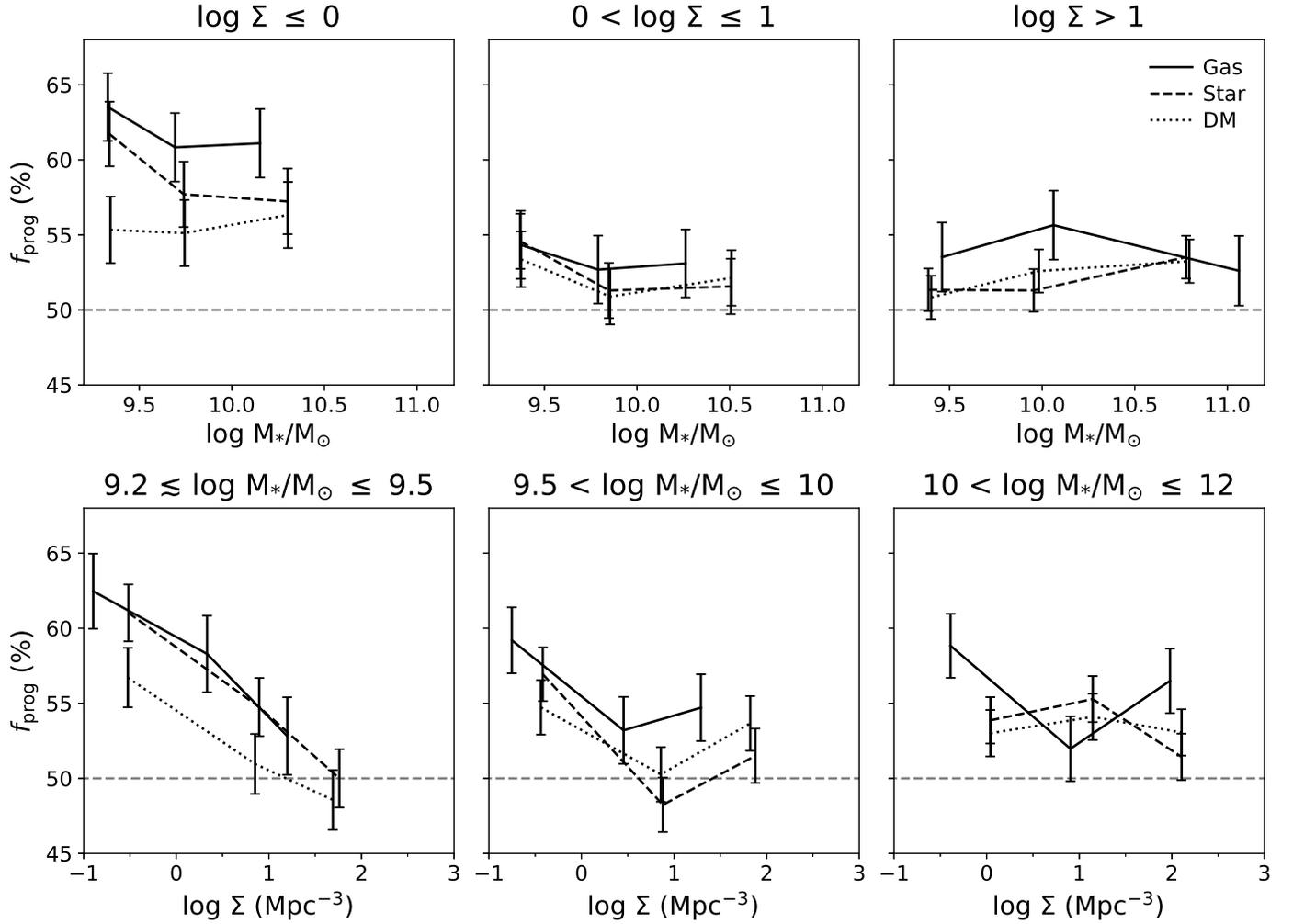}
\caption{Top: prograde fraction of the spin--orbit angle as a function of the stellar mass within fixed ranges of the local density. The spin--orbit angle of each galaxy is calculated separately for gas (solid lines), stars (dashed lines), and DM (dotted lines). For each line, the three bins contain an equal number of galaxies. Error bars represent the standard error of the percentage. Bottom: the same as the upper panels, but as a function of the local density within fixed ranges of the stellar mass. \label{fig:fprogme}}
\end{figure*}

Figure~\ref{fig:fprogme} presents the prograde fraction as functions of the stellar mass and the local density. As shown in the top panels, the dependence on the stellar mass is marginal. The prograde fraction for stars (solid lines) increases with decreasing the stellar mass only when the galaxies reside in low-density environments (log $\Sigma$ $\leq$ 0), but the fraction for gas (dashed lines) and DM (dotted lines) or for galaxies in dense environments (log $\Sigma$ $>$ 1) does not depend on the stellar mass. On the other hand, the dependence on the local density in the bottom panels is more pronounced. Paired galaxies in low-density regions generally show a higher prograde fraction than those in high-density regions. This trend is valid for all components (gas, stars, and DM) of low- and intermediate-mass target galaxies ($M_\mathrm{*}$ $\leq$ 10$^{10}$\,$M_\odot$). To summarize, the SOA signal becomes more robust as both the stellar mass and the local density decrease, and the local density seems to be a more critical factor than the stellar mass. \citet{2019ApJ...872...78L} observed that faint or blue target galaxies have a stronger alignment than bright or red ones. The trend we find is consistent with theirs, given the fact that low-density regions are largely populated by blue galaxies \citep[e.g.,][]{2009MNRAS.393.1324B}, although they did not directly inspect the local environment.

\begin{figure*}[tb!]
\epsscale{1.17}
\plotone{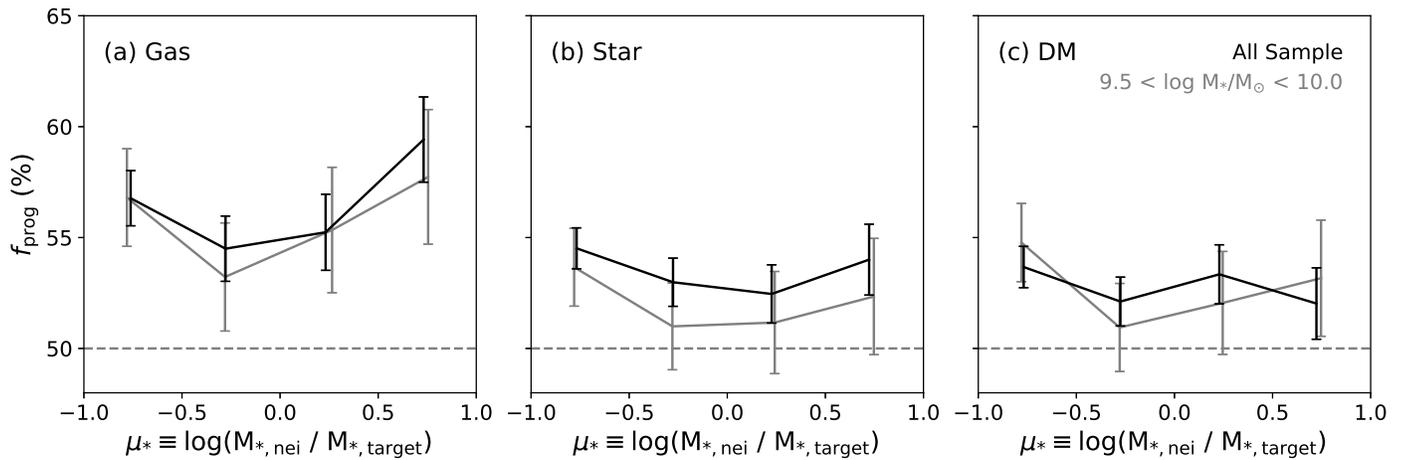}
\caption{Prograde fraction of the spin--orbit angle calculated for (a) gas, (b) stars, and (c) DM as a function of the stellar mass ratio of the neighbor to the target. Black lines are for the whole sample, and gray lines are for target galaxies with 10$^{9.5}$ $<$ $M_\mathrm{*}/M_\odot$ $<$ 10$^{10}$. Error bars represent the standard error of the percentage. \label{fig:fprogmu}}
\end{figure*}

Figure~\ref{fig:fprogmu} shows the prograde fraction as a function of the stellar mass ratio between the target and the nearest neighbor. We do not find any significant effect of the stellar mass ratio, but there is a slight increase in the prograde fraction for pairs with large $|\mu_*|$ compared to those with small $|\mu_*|$. Gray lines show the case for target galaxies with 10$^{9.5}$ $<$ $M_\mathrm{*}/M_\odot$ $<$ 10$^{10}$ to get rid of the effect of stellar mass, but the result is barely different from the whole sample. The dependence of the alignment on the mass ratio is generally insignificant within the range of $1/10 < \mu_* < 10$, and all results of this paper are qualitatively unchanged if we restrict our analysis only to equal-mass pairs (e.g., $1/3 < \mu_* < 3$).

\subsection{Impact of Halo-scale Environments} \label{subsec:dependhalo}

\begin{figure*}[tb!]
\epsscale{1.17}
\plotone{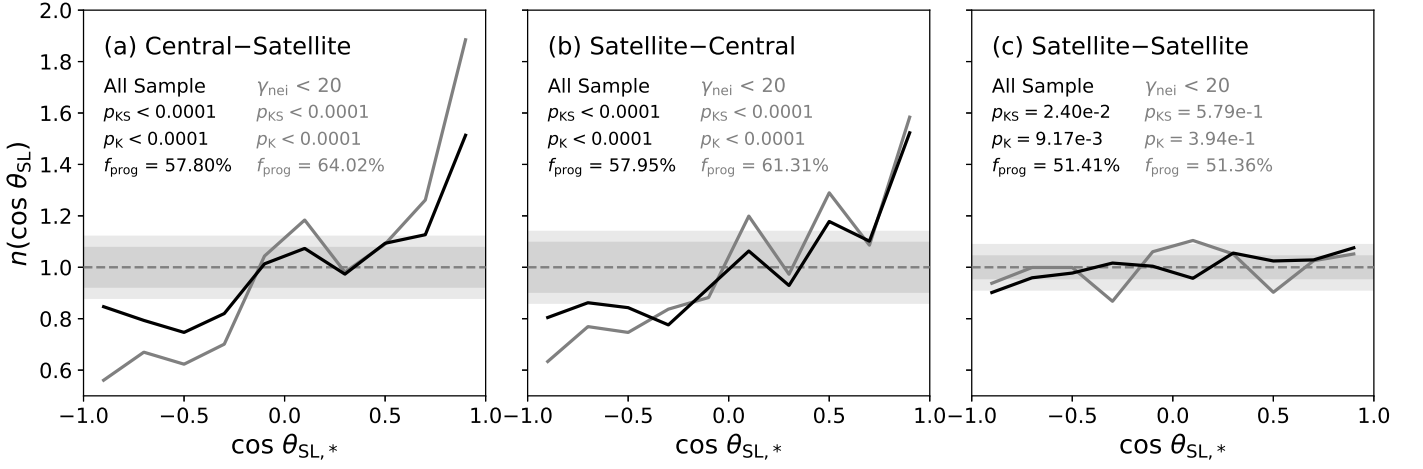}
\caption{PDF of the stellar spin--orbit angle of (a) central--satellite pairs (central targets with a satellite neighbor), (b) satellite--central pairs (satellite targets with a central neighbor), and (c) satellite--satellite pairs (satellite targets with a satellite neighbor). Black lines are for the whole sample, and gray lines are for paired galaxies with $\gamma_\mathrm{nei}$ $<$ 20. The format is the same as in Figure~\ref{fig:soa}. \label{fig:soacensat}}
\end{figure*}

\begin{figure*}[tb!]
\epsscale{1.17}
\vspace{-0.7mm}
\plotone{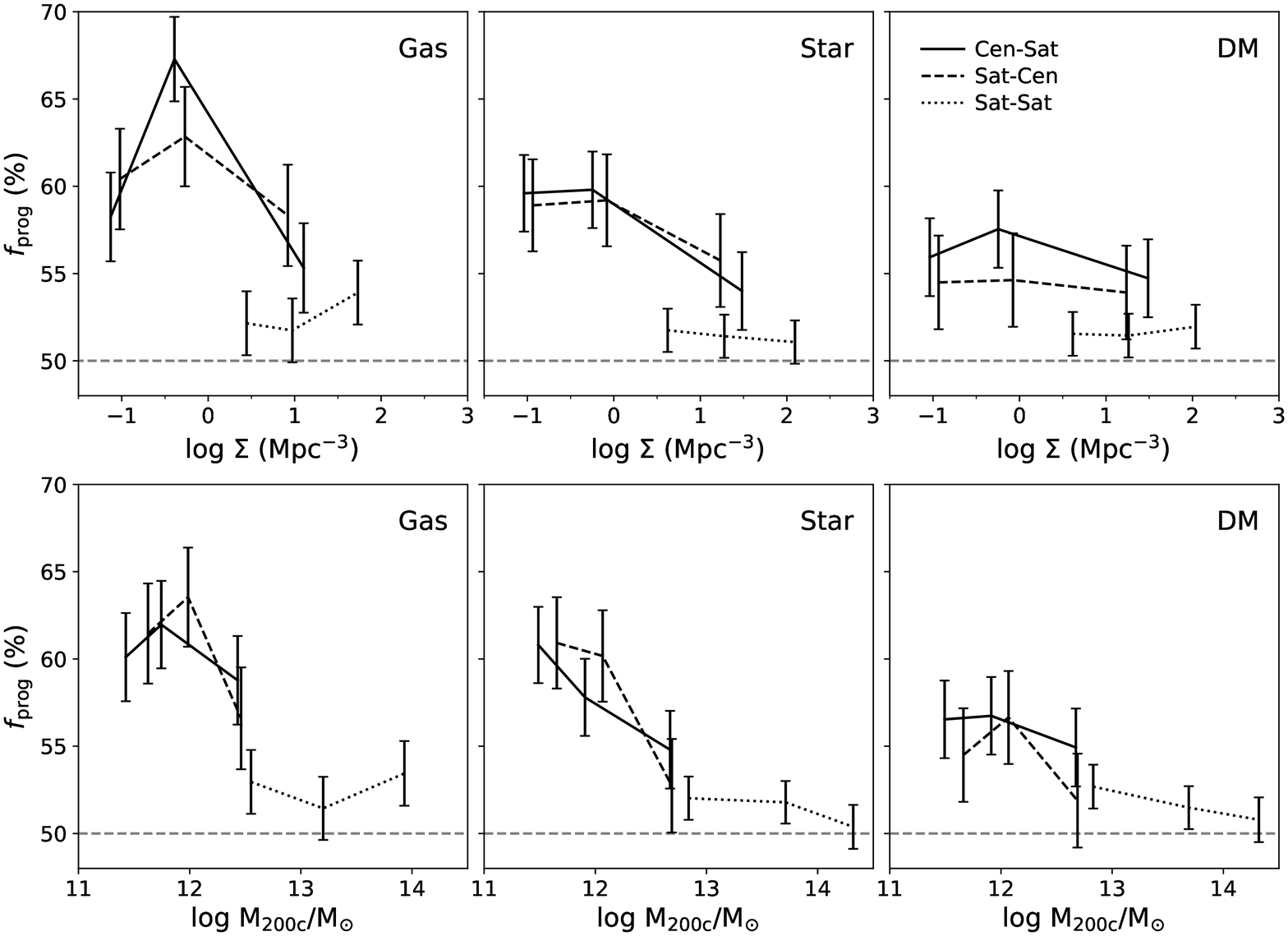}
\caption{Top: prograde fraction of the spin--orbit angle as a function of the local density for central--satellite pairs (solid lines), satellite--central pairs (dashed lines), and satellite--satellite pairs (dotted lines). The spin--orbit angle of each galaxy is calculated separately for gas (left), stars (middle), and DM (right). For each line, the three bins contain an equal number of galaxies. Error bars represent the standard error of the percentage. Bottom: the same as the top panels, but as a function of the halo mass (the mass enclosed in a sphere with a mean density of 200 times the critical density). \label{fig:fprogcensat}}
\end{figure*}

The presence of the SOA governed by environment raises additional questions: Does the halo-scale environment play a particular role in the formation of the SOA? Does it matter if a galaxy is a central or a satellite in a halo? Is the halo-scale environment more important than the large-scale one? We try to answer such questions in this subsection. Figure~\ref{fig:soacensat} shows the PDF of the stellar spin--orbit angle for subsamples split by whether the target and neighbor are a central or a satellite. Since the target and neighbor in each pair share the same FoF halo by definition (see Section~\ref{subsec:sample}), there are three possible target--neighbor combinations: central--satellite, satellite--central, and satellite--satellite pairs. We can see that there is a clear difference between the subsamples. The central--satellite pairs in panel (a) exhibit a clear signal of SOA; the $p$-values of both the KS and Kuiper's tests are $<$ 0.0001, and the prograde fraction is 57.8 $\pm$ 1.28\,\%. Also, the subsample of the satellite--central pairs in panel (b) shows the alignment, and the strength of the alignment is as high as that of the central--satellite pairs; the $p$-values of both tests are $<$ 0.0001, and the prograde fraction is 58.0 $\pm$ 1.53\,\%. However, unlike the former two groups, the satellite--satellite pairs in panel (c) do not show a significant level of SOA; $p_\mathrm{KS}$ is only 0.024, $p_\mathrm{K}$ is 0.009, and the prograde fraction is 51.4 $\pm$ 0.71\,\%, which is close to isotropic distribution. This result is unchanged if we restrict the sample to close pairs with $\gamma_\mathrm{nei} < 20$ (shown as gray lines). While the prograde fraction for the central--satellite and the satellite--central pairs increases with decreasing the separation as already shown in Figure~\ref{fig:fprogdist}, the satellite--satellite pairs alone show no signal of alignment ($p_\mathrm{KS}$ $=$ 0.58, $p_\mathrm{K}$ $=$ 0.39, and $f_\mathrm{prog}$ $=$ 51.4 $\pm$ 1.48\,\%), even at a small separation ($\gamma_\mathrm{nei} < 20$).

It is obvious that the SOA is developed by interactions between a central and its satellites. This reminds us of the so-called radial alignment in galaxy groups. It is known that the major axes of satellites in a massive group tend to be aligned toward the direction of the central galaxy, which is due to the tidal torque induced by the DM halo embedding the satellites \citep[e.g.,][]{2005ApJ...627L..21P, 2008ApJ...672..825P, 2015MNRAS.451.2536R}. Given the radial alignment, it is possible to expect that the spin of the satellites is (perpendicularly) aligned with the direction to the central because the tidal torque also changes the angular momentum perpendicular to the orbital plane \citep{1997ApJ...487..489U}. However, the main difference between the radial alignment and the SOA is that the latter is more prevalent in low-density regions (see Figures~\ref{fig:soame} and \ref{fig:fprogme}), while the former is explained by the assumption of a massive group which is abundant in dense environments.

In Figure~\ref{fig:fprogcensat}, we investigate the correlation between the SOA and the inter- and intra-halo-scale environments. The top panels show the prograde fraction as a function of the local density for central--satellite pairs (solid lines), satellite--central pairs (dashed lines), and satellite--satellite pairs (dotted lines). Interestingly, there is a strong connection between the central/satellite type and the local density. The satellite--satellite pairs are in environments distinct from the others in that the majority of them reside in dense regions. This seems reasonable because the number ratio of satellites to centrals is high in dense regions such as clusters or groups, and massive centrals in such regions are also difficult to meet the mass ratio criterion (i.e., $|\mu_*| < 1.0$) with much less massive satellites. Therefore, the absence of the SOA for the satellite--satellite pairs in Figures~\ref{fig:soacensat} and \ref{fig:fprogcensat} is essentially identical to the weak alignment in high-density regions shown in Figures~\ref{fig:soame} and \ref{fig:fprogme}. Nevertheless, central--satellite (and satellite--central) pairs show a higher prograde fraction than satellite--satellite pairs at fixed local density (e.g., 0.5 $<$ log $\Sigma$ $<$ 1.5). Besides, the dependence of the prograde fraction on the local density within each subgroup in Figure~\ref{fig:fprogcensat} is weakened compared to that of Figure~\ref{fig:fprogme}. This result indicates the importance of the central/satellite type over the local density. 

In the bottom panels of Figure~\ref{fig:fprogcensat}, we now plot the prograde fraction as a function of the FoF halo mass ($M_{200c}$), defined as the total mass enclosed within a sphere with a mean density equal to 200 times the critical density. The prograde fraction decreases as the halo mass increases. As expected, the majority of satellite--satellite pairs are located in massive halos, and the difference in the mean halo mass between the satellite--satellite and central--satellite pairs is more dramatic than that in the local density. A significant SOA is evident only for galaxies in less massive halos, and hence mostly central--satellite (and satellite--central) pairs. 

In this paper, we do not attempt to disentangle further the relative importance among the local density, the halo mass, and the central/satellite type due to their strong degeneracy. The message from Figure~\ref{fig:fprogcensat}, however, is quite clear; the SOA is better developed in simpler systems where only a few galaxies are involved. We may infer from this that the SOA is created by long-lasting interactions. Galaxy pairs in massive groups (mostly satellite--satellite pairs) are often unbound to each other and influenced by a third neighbor \citep{2013MNRAS.436.1765M}, and therefore they are predominantly in flyby interactions \citep{2019ApJ...887...59A}. In contrast, pairs in low-density regions and less massive halos are likely to interact with a single neighbor for an extended period of time.

\section{Discussion} \label{sec:discussion}

Our analysis shows that the spin direction of each galaxy in a pair is in alignment with the orbital angular momentum vector of the pair system. The alignment is stronger as the separation to the neighbor decreases (Section \ref{subsec:soatng}). The signal increases weakly as the target's stellar mass decreases (Section \ref{subsec:dependme}). The strength of the alignment is closely related to the environment where the galaxy pairs reside, showing a significant signal at low-density regions (Section \ref{subsec:dependme}) and for interactions between centrals and satellites in low-mass halos (Section \ref{subsec:dependhalo}). 

The spin direction of galaxies is often associated with the geometry of the LSS. One can anticipate that the anisotropic nature of the LSS helps to generate the SOA in galaxy pairs. However, the general trend of the SOA is not fully consistent with this picture. For instance, the spin--LSS alignment appears stronger for massive galaxies located in filaments \citep[e.g.,][]{2018MNRAS.481.4753C,  2019MNRAS.487.1607G, 2020MNRAS.493..362K}, while the SOA is stronger for less massive galaxies in sparser environments. For galaxy pairs in filaments, the orbital plane's normal vectors tend to be perpendicular to the filament axis \citep{2015A&A...576L...5T, 2018A&A...619A..24M}, which seems to conflict with the finding that less massive galaxies, showing a stronger signal of the SOA, prefer spin vectors parallel to the filaments \citep[e.g.,][]{2018ApJ...866..138W, 2018A&A...613A...4W, 2020MNRAS.491.2864W}. Besides, the preferred direction of satellite accretion along the cosmic flow is expected to be more pronounced in massive halos \citep{2014MNRAS.443.1274L, 2015ApJ...813....6K, 2015MNRAS.450.2727T}, but the SOA is more significant in less massive halos.

\begin{figure*}[tb!]
\epsscale{1.17}
\plotone{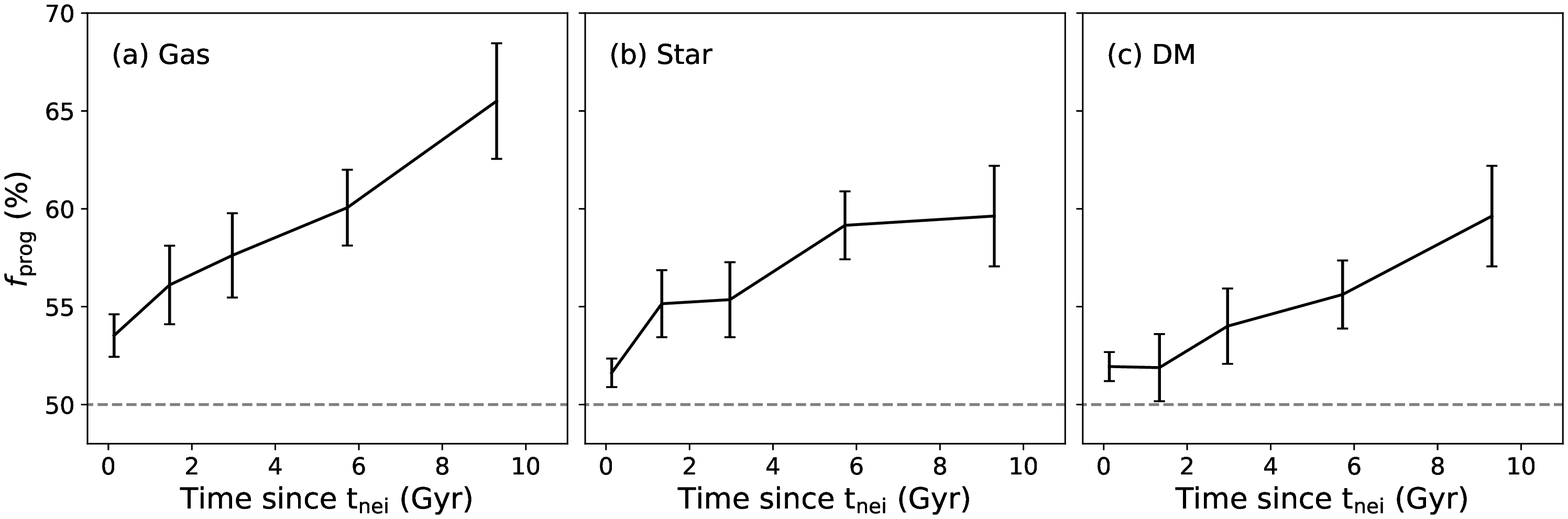}
\caption{Prograde fraction of the spin--orbit angle calculated for (a) gas, (b) stars, and (c) DM as a function of $t_{\mathrm{nei}}$ (see the definition in Section \ref{subsec:pairedgalaxies})---that is, the time span of the interaction after the current nearest neighbor remains the same without change. Error bars represent the standard error of the percentage. \label{fig:fprogtnei}}
\end{figure*}

The properties of the SOA indicate that the formation of the alignment is related to long-term interactions between paired galaxies rather than set by the underlying LSS. Figure~\ref{fig:fprogtnei} shows the correlation between the prograde fraction and the time when a galaxy is paired with its current neighbor ($t_{\mathrm{nei}}$; see the definition in Section \ref{subsec:pairedgalaxies}). An earlier $t_{\mathrm{nei}}$ means that the galaxy interacts with its current neighbor for a more extended period without the interference of a third neighbor. The figure shows a strong correlation between the prograde fraction and the duration of the interaction. Galaxy pairs interacting for a longer time generally show a higher prograde fraction compared to recently formed pairs for gas, stars, and DM in common. While the study on the effect of interactions on the galaxy spin mainly focused on its strength but not the direction \citep[e.g.,][]{2010MNRAS.402.1807C, 2018ApJ...856..114C, 2018MNRAS.477.1567L}, some recent studies hinted that the interactions could develop the dynamical alignment of galaxy pairs \citep[e.g.,][]{2019ApJ...872...78L, 2020ApJ...893..154L}. In particular, \citet{2020ApJ...893..154L} observed that paired galaxies with an SOA-like coherence also exhibit similarity in colors and claimed that this is evidence of recent interactions between the two galaxies in alignment. The results in this paper are compatible with this interaction scenario. The interaction can easily influence low-mass galaxies with low angular momenta (see Figures~\ref{fig:soame} and \ref{fig:fprogme}). A more significant SOA is developed for pairs with an earlier $t_{\mathrm{nei}}$ (see Figure~\ref{fig:fprogtnei}), which on average have a closer neighbor (see Figures~\ref{fig:soadist} and \ref{fig:fprogdist}) and reside in less massive halos at lower-density regions (see Figure~\ref{fig:fprogcensat}).

The spin and shape of a satellite are known to be regulated by the tidal torque induced by the massive central halo (a. k. a., the radial alignment of satellites) and has been investigated in many previous studies \citep[e.g.,][]{1997ApJ...487..489U, 2005ApJ...627L..21P, 2008ApJ...672..825P, 2015MNRAS.451.2536R}. Our results are in line with these studies, especially in the sense that the alignment is created between centrals and satellites (see Figure~\ref{fig:soacensat}), but the main difference is that the SOA signal is dominated by low-mass halos (see Figure~\ref{fig:fprogcensat}). While the difference is likely because we restrict our sample to pairs of comparable mass, a question remains whether the tidal torque of the comparable-mass neighbor is sufficient to change the spin of the target. Moreover, it is counterintuitive that there is no clear trend of the SOA with the mass ratio (see Figure~\ref{fig:fprogmu}). Some theoretical studies predict that the radial alignment can be generated in low-mass halos less massive than the Milky Way--sized ones \citep[e.g.,][]{2008ApJ...672..825P, 2020MNRAS.495.3002K}, but more detailed studies are needed.

In addition to the tidal effect, direct collisions between gas in contact pairs may facilitate the formation of the SOA. For example, hydrodynamic forces such as ram-pressure can cause an abrupt change in the spin of the gas component during close encounters \citep[e.g.,][]{2017MNRAS.465.2643C}. In this scenario, the alignment is first established for the gas, and then stars born therein inherit the angular momentum from the gas, while the DM is only weakly associated with the others through gravitational forces. This can explain the reason why the SOA signal decreases in the order of gas, stars, and DM (see Figure~\ref{fig:soa}). The ram-pressure is usually more violent in retrograde interactions \citep{2002MNRAS.333..481B, 2018MNRAS.479.3952B}, and the selective removal of retrograde systems could be partially responsible for the high prograde fraction.
% Furthermore, when a galaxy moves through a rotating structure of its neighbor, dynamical friction exerts a force opposite to the relative motion, reducing the orbital angular momentum anti-parallel to the spin vector of the neighbor. This process was originally proposed to account for the pairing of BHs in retrograde galaxy mergers \citep[e.g.,][]{2020MNRAS.494.3053B}, but dynamical friction partially contributes to the formation of alignments for paired galaxies \citep{1991ApJ...375L...5K, 2000A&A...355..532W} and orbiting satellites \citep{2002MNRAS.333..779P, 2004MNRAS.349..747P}.

The observation of the SOA is rather challenging because the dynamics of both galaxies in pairs should be determined. There are very few observations to be compared to our simulation results. \citet{2019ApJ...872...78L} observed the SOA-like coherence up to a distance of 800\,kpc, which is similar to the spatial scale we found. Their findings are compatible with many of our results (e.g., the trend with mass and color) but not all (e.g., the trend with mass ratio). However, a direct quantitative comparison is difficult due to the difference in the sample selection and methodology. Specifically, \citet{2019ApJ...872...78L} selected multiple neighbors for each target and inspected the averaged (projected) radial velocity. Although it would be possible to generate a mock observation for a better comparison, it is not trivial and beyond the scope of this paper. Several observational studies have focused on the orbital motion of small satellites within a group, rather than galaxy pairs, generally showing no clear preference \citep{1997ApJ...478L..53Z, 2008MNRAS.384..803H, 2010ApJ...720..522H}. This is understandable since our results also do not show strong evidence for the SOA in massive halos. Obviously, more observations are required to confirm the predictions of this paper. We expect that ongoing IFU surveys would be useful to perform more observational tests in the near future, considering that these surveys contain a number of galaxy pairs with kinematic data available \citep[e.g.,][]{2015A&A...582A..21B, 2020ApJ...892L..20F}.

We leave a more in-depth investigation of the mechanisms behind the SOA to future work. In this study, using a cosmological hydrodynamic simulation, we investigated the SOA of galaxy pairs only at $z = 0$ and provided significant evidence for the SOA. Also, we studied its dependence on the mass and environment and concluded that the current result favors the interaction origin. However, this analysis alone is not conclusive as to the origin of the alignment. To further understand the origin of the SOA, a detailed investigation of the formation and evolution of the SOA over cosmic time is required. Specifically, this includes the key question of the timing of the SOA formation. For instance, we may expect this: if the SOA is created by the primordial tidal torquing, the alignment occurs in the very early universe; if the SOA is developed by the preferential accretion, it is established before the galaxy is paired with its current neighbor; if the SOA is related to the tidal force from the neighbor, it is tied to the orbital phase of the system; or if the SOA is produced by the contact encounter, it forms only after the close passage. We plan to clarify this issue and further constrain the underlying physical mechanism for the alignment in forthcoming papers. 

\section{Conclusion} \label{sec:conclusion}

We have analyzed the alignment between the spin vector of a target and the orbital angular momentum vector of its nearest neighbor, which we refer to as the spin--orbit alignment (SOA). We used a cosmological hydrodynamic simulation of the IllustrisTNG project (Section~\ref{subsec:simul}) and identified paired galaxies with comparable-mass neighbors at $z = 0$ (Section~\ref{subsec:sample}). Our final sample consists of 7434 target galaxies with the updated mass and size, which are corrected for the artificial mass loss of satellites (Section~\ref{subsec:pairedgalaxies}). We calculated the angle between the spin and the orbital angular momentum vectors (the spin--orbit angle; \thetasl{}) of the pairs separately for each component (i.e., gas, star, and DM; Section~\ref{subsec:spin-orbit}). We have found a significant preference for prograde orientations ($|$\thetasl{}$|$ $<$ 90$^{\circ}$) for galaxy pairs at $z = 0$. We investigated the dependence of the SOA on galaxy properties, including mass and environment (Section~\ref{sec:results}). Our results are summarized as follows:
\begin{enumerate}
\setlength\itemsep{0em}
\item Galaxy pairs at $z = 0$ statistically prefer prograde orientations ($|$\thetasl{}$|$ $<$ 90$^{\circ}$). The significance over the isotropic sample at \costheta{} $\sim$ 1 is 9.8$\sigma$ for gas, 6.5$\sigma$ for stars, and 4.9$\sigma$ for DM. The alignment is the strongest for gas and the weakest for DM (Section~\ref{subsec:soatng}).
\item The strength of the SOA increases with decreasing separation from the nearest neighbor. The number fraction of pairs in prograde orientation is as high as 70\,\% for gas spins and 60\,\% for stars and DM at the separation of $\gamma_\mathrm{nei} < 10$. The SOA signal extends out to $\gamma_\mathrm{nei} \sim 100$, which is roughly 600\,kpc on physical scale (Section~\ref{subsec:soatng}).
\item Galaxy pairs in lower-density environments show a higher prograde fraction, which holds true for low- and intermediate-mass target galaxies ($M_\mathrm{*}$ $\leq$ 10$^{10}$\,$M_\odot$). The prograde fraction slightly increases for low-mass galaxies, but the dependence on the stellar mass is less pronounced compared to the dependence on the local density. The spin--orbit angle does not show a clear trend with the mass ratio of interacting pairs (Section~\ref{subsec:dependme}).
\item The SOA is created only between centrals and satellites. A clear signal is detected in both central--satellite and satellite--central pairs. The central/satellite type is closely linked to the local density and the halo mass. Galaxy pairs with a strong SOA are mostly located in low-density regions and low-mass halos (Section~\ref{subsec:dependhalo}).
\item The prograde fraction increases with the interaction duration with the current neighbor ($t_{\mathrm{nei}}$). The SOA can be created either by the effect of galaxy--galaxy interactions or by the anisotropic nature of the LSS. The current evidence favors the scenario where the SOA is developed by interactions with a neighbor for an extended period without the interference of a third neighbor, rather than the influence of the underlying LSS (Section~\ref{sec:discussion}).
\end{enumerate}

\acknowledgments This work is supported by the Mid-career Researcher Program (No. 2019R1A2C3006242) and the SRC Program (the Center for Galaxy Evolution Research; No. 2017R1A5A1070354) through the National Research Foundation of Korea.

\bibliography{SOATNG}{}
\bibliographystyle{aasjournal}

%% This command is needed to show the entire author+affiliation list when
%% the collaboration and author truncation commands are used.  It has to
%% go at the end of the manuscript.
%\allauthors

%% Include this line if you are using the \added, \replaced, \deleted
%% commands to see a summary list of all changes at the end of the article.
%\listofchanges

\end{document}